\documentclass{iopart}
\usepackage[T1]{fontenc}
\usepackage[latin9]{inputenc}
\usepackage{units}
\usepackage{amstext}
\usepackage{amssymb}
\usepackage{esint}

\makeatletter
\usepackage{iopams}
\usepackage{setstack}

\makeatother

\begin{document}

\title{Resources needed for non-unitary quantum operations}

\author{Raam Uzdin}

\address{Institute for Theoretical Atomic, Molecular and Optical Physics,
Harvard-Smithsonian Center for Astrophysics}

\address{Present address: Racah Institute of Physics, The Hebrew University,
Jerusalem, Israel.}

\ead{raam@mail.huji.ac.il}
\begin{abstract}
Non-unitary operations generated by an effective non-Hermitian Hamiltonian
can be used to create quantum state manipulations which are impossible
in Hermitian quantum mechanics. These operations include state preparation
(or cooling) and non-orthogonal state discrimination. In this work
we put a lower bound on the resources needed for the construction
of some given non-unitary evolution. Passive systems are studied in
detail and a general feature of such a system is derived. After interpreting
our results using the singular value decomposition, several examples
are studied analytically. In particular, we put a lower bound on the
resources needed for non-Hermitian state preparation and non-orthogonal
state discrimination. 
\end{abstract}
\maketitle
\global\long\def\ket#1{\left|#1\right\rangle }

\global\long\def\bra#1{\left\langle #1\right|}

\global\long\def\braket#1#2{\left\langle #1|#2\right\rangle }

\global\long\def\ketbra#1#2{\left|#1\right\rangle \left\langle #2\right|}

\global\long\def\braOket#1#2#3{\left\langle #1\left|#2\right|#3\right\rangle }

\global\long\def\mc#1{\mathcal{#1}}

\global\long\def\nrm#1{\left\Vert #1\right\Vert }

\section{Introduction}

Unitary quantum evolution generated by Hermitian Hamiltonian conserves
the angle between states. For example, two states which are orthogonal
at one instant will remain orthogonal throughout the evolution. In
effective models of some sub-systems the Hamiltonian may be non-Hermitian.
This may be either due to leakage of particles outside the region
of interest, (as happens, for example, in Stark ionization), or due
to absorption or gain, as often occurs in optics. Another source of
non-unitarity can appear when some measurements are performed on entangled
systems (e.g \cite{NH entang meas}). For additional information see
\cite{Nimbook}. Effective non-Hermitian Hamiltonians lead to non-unitary
evolution operators. It was suggested in \cite{PT dis} that non-Hermitian
Hamiltonian can be used for perfect discrimination of two non-orthogonal
states. In principle this scheme can work even if the states are almost
perfectly parallel to each other. In \cite{PT dis} the question of
the needed resources for this task was not addressed. In this work,
we have fixed a lower bound on the resources needed for a non-unitary
task in order to address this question.

We start in Sec. \ref{sec:Quantifying-evolution-resources} by introducing
the ``norm action'' as a plausible measure for the resources needed
to generate some evolution. In Sec. \ref{sec:A-lower-bound BASIC}
we derive our basic result: a lower bound on the norm action that
depends only on the evolution operator at the end of the process.
That is, a simple bound can be set on the desired evolution resources
without knowing anything on the Hamiltonian. Passive systems are discussed
in sec. \ref{sec:Improved and passive} where our basic result is
generalized so that it will be effective for passive systems as well.
In Sec. \ref{sec:purely-non-unitary-evolution} we use the singular
value decomposition to explain in what sense our bound quantifies
solely the ``purely non-unitary'' part of the evolution. Next, in
section \ref{sec:Examples} we analytically study several examples
in order to illustrate various points concerning our main result.
Section \ref{sec:Concluding-remarks} contains concluding remarks
and the three appendices contain some clarification.

\section{Quantifying evolution resources\label{sec:Quantifying-evolution-resources}}

Let $U(t)\in\mathbb{C}^{N\times N}$ be some some non-unitary evolution
operator in an $N$-level quantum system. The action of $U(t)$ on
a state $\ket{\psi(t=0)}$ is: 
\begin{equation}
\ket{\psi(t)}=U(t)\ket{\psi(0)}.
\end{equation}
In general, the determinant of $U(t)$ is not fixed in time, and it
satisfies the Liouville formula \cite{Liou formula}:
\begin{equation}
\det[U(t)]=\mbox{det}[U(0)]e^{-i\intop_{0}^{t}\mbox{tr}[H(t')]dt'},
\end{equation}
where $H$ is the Hamiltonian of the system. The trace of $H$ only
contributes a complex multiplier to the evolution operator. Obviously
the trace affects the required resources, but these resources are
rather trivial as they amount to adding a common phase or attenuation
to all states. In the first stage we are only interested in resources
associated with actual state transformation. Therefore, to avoid the
trivial changes in $U$ generated by the trace of $H$ we define the
traceless Hamiltonian: 
\begin{equation}
\mc H=H-\frac{\mbox{tr}(H)}{N}I.
\end{equation}
The evolution operator $\mathcal{U}$ generated by this Hamiltonian
satisfies:
\begin{equation}
\mbox{det}[\mathcal{U}(t)]=1,\label{eq: det norm}
\end{equation}
and it is related to $U$ through: 
\begin{equation}
\mc U=U/det[U]{}^{1/N}\label{eq: cal U def}
\end{equation}
$\mc U$ is the determinant normalized version of $U$. The geometric
meaning of condition (\ref{eq: det norm}) is given in appendix I.
Despite (\ref{eq: det norm}), note that $\mathcal{U}$ is in general
not a unitary operator:
\begin{equation}
\mathcal{U}^{\dagger}\mathcal{U}\neq\mbox{I},
\end{equation}
where $\dagger$ stands for the complex transposition operation. Assume
there is some non-Hermitian Hamiltonian $H(t)$ that after some time
$T$ generates a desired evolution $U(T)$. Our goal will be to use
$U(T)$ to set limits on the smallness of $H$. However, a simple
scaling argument shows that for any final $U$, the generating Hamiltonian
elements can be made arbitrarily small by making the evolution slower.
To remove this ambiguity we will examine the time integral over the
Hamiltonian ``size``. The scalar quantity we shall use to quantify
the Hamiltonian size at each instant is the matrix norm of the Hamiltonian
matrix. In summary, our goal is to set a lower bound on the action-like
quantity:
\begin{equation}
\int_{0}^{T}\nrm{H(t)}dt\label{eq: norm action}
\end{equation}
We shall refer to this quantity as the ``norm action''. In this
work $\nrm{\cdot}$ refers to any matrix norm that satisfies the sub-multiplicative
property:
\begin{equation}
\nrm{AB}\le\nrm A\nrm B,\label{eq: sub mul}
\end{equation}
In particular, we shall often use the spectral norm \cite{horn} of
the Hamiltonian. A previous study \cite{100=000025 efficiency} showed
that the spectral norm of $\mc H$ constitutes an upper bound on the
evolution speed%
\footnote{Loosely speaking the angular speed in ``Bloch Sphere''%
} of a quantum state driven by NH or Hermitian Hamiltonian. This physical
meaning of the spectral norm makes it our default choice of norm.
Nonetheless, some of the results presented here equally apply to any
sub-multiplicative norm just as well. Therefore the spectral norm
will be used only where it can provide some additional insights unique
to this norm. The (sub-multiplicative) spectral norm of the Hamiltonian
is the largest singular value of $H$ \cite{horn 2} and it is equal
to:
\begin{equation}
\nrm H_{s}=\sqrt{\max[\mbox{eig}\{H^{\dagger}H\}]},\label{eq: spec norm def}
\end{equation}
where $\mbox{eig}\{\}$ is just the list of eigenvalues of the enclosed
operator. Before we proceed, it is worth mentioning two properties
of the spectral norm which provide more explicit information on the
Hamiltonian matrix elements. Let $H_{i,j}$ be the elements of an
operator $H\in\mathbb{C}^{N\times N}$. The spectral norm satisfies:

\begin{eqnarray}
\nrm{H(t)}_{s} & \le & \sqrt{\sum_{i,j=1}^{N}\left|H_{i,j}\right|^{2}}\label{eq: spec HS}\\
\max\left|H_{i,j}\right|\le\nrm{H(t)}_{s} & \le N & \max\left|H_{i,j}\right|\label{eq: spec max}
\end{eqnarray}
That is, a lower bound on (\ref{eq: norm action}) would also constitute
a lower bound on $\intop_{0}^{T}\sqrt{\sum_{i,j=1}^{N}\left|H_{i,j}(t)\right|^{2}}dt$
and on $\intop_{0}^{T}\max\left|H_{i,j}(t)\right|dt$.

\section{A lower bound on the resources needed for non-unitary evolution\label{sec:A-lower-bound BASIC}}

In this section we derive a lower bound on the ``norm action'' (\ref{eq: norm action}).
The derivation is carried for the more general $H$ and $U$ and only
later replace them by $\mc H$ and $\mc U$. The evolution operator
$U(t)$ satisfies:
\begin{equation}
i\frac{d}{dt}U=HU\label{eq: schrod U}
\end{equation}
Taking the norm and using sub-multiplicativity (\ref{eq: sub mul}):
\begin{equation}
\nrm{\frac{d}{dt}U}=\nrm{HU}\le\nrm H\nrm U,
\end{equation}
or:

\begin{equation}
\nrm H\ge\nrm U^{-1}\nrm{\frac{d}{dt}U}
\end{equation}
Since any norm satisfies the ``inverse triangle inequality'': $\nrm{A-B}\ge\left|\nrm A-\nrm B\right|$,
we get:

\begin{eqnarray}
\nrm{\frac{d}{dt}U} & = & \lim_{dt\to0}\frac{\nrm{U(t+dt)-U(t)}}{dt}\nonumber \\
 & \ge & \left|\frac{\nrm{U(t+dt)}-\nrm{U(t)}}{dt}\right|=\left|\frac{d}{dt}\nrm{U(t)}\right|
\end{eqnarray}
Therefore:
\begin{equation}
\nrm H\ge\nrm U^{-1}\left|\frac{d}{dt}\nrm{U(t)}\right|\ge\nrm U^{-1}\frac{d}{dt}\nrm{U(t)}
\end{equation}
Integrating we get our basic resources bound:

\begin{equation}
\int_{0}^{T}\nrm{H(t)}dt\ge\ln\left\Vert U(T)\right\Vert \label{eq: basic ineq with trace}
\end{equation}
An alternative derivation is given in appendix II. 

Although this equation is correct, it is sensitive to $\mbox{tr}(H)$
and $\det[U]$ which are related to the somewhat irrelevant information
discussed earlier. This can be easily resolved by noticing that $\mc H$
and $\mc U$ also satisfy $i\frac{d}{dt}\mc U=\mc{HU}$, so (\ref{eq: basic ineq with trace})
can be used to obtain: 

\begin{equation}
\int_{0}^{T}\nrm{\mc H(t)}dt\ge\ln\left\Vert \mathcal{U}(T)\right\Vert .\label{eq: basic ineq}
\end{equation}
One may be concerned that perhaps $\left\Vert \mathcal{U}(T)\right\Vert \le1$
and therefore the RHS of (\ref{eq: basic ineq}) is negative which
makes Eq. (\ref{eq: basic ineq}) trivial and useless. Nonetheless,
for the spectral norm, one can show that:
\begin{equation}
\left\Vert \mathcal{U}(T)\right\Vert _{s}=1\Leftrightarrow\mc{U\mbox{ is unitary}}\label{eq: spec of unitary}
\end{equation}
Thus, for non-unitary operators the RHS of (\ref{eq: basic ineq})
is larger than zero and our lower bound is non-trivial. A simple derivation
(\ref{eq: spec of unitary}) is given in appendix III. Using the following
property of the spectral norm:
\begin{equation}
\nrm A_{s}\ge\max\left|\mbox{eig}\{A\}\right|\label{eq: spect radius}
\end{equation}
For any operator $A\in\mathbb{C}^{N\times N}$, it is possible to
obtain a weaker lower bound in terms of the eigenvalues of evolution
operator:

\begin{equation}
\int_{0}^{T}\nrm{\mc H(t)}_{s}dt\ge\ln[\max\left|\mbox{eig}\{\mc U\}\right|].\label{eq: eig ineq}
\end{equation}
In our third example we will examine a non-diagonalizable case (NH
degeneracy) where all the eigenvalues are equal to one so that the
RHS of (\ref{eq: eig ineq}) is equal to zero. Nonetheless, the operator
is still non-unitary and the stronger resources lower bound (\ref{eq: basic ineq})
is greater than zero.

The fact that the RHS of (\ref{eq: basic ineq}) is equal to zero
for unitary evolution operators (due to (\ref{eq: basic ineq}) and
(\ref{eq: spec of unitary})) gives some indication that our spectral
norm bound quantifies the resources needed for non-unitary evolution
but not the resources for unitary evolution. This point will be clarified
in Sec. \ref{sec:purely-non-unitary-evolution}. However, before doing
that we wish to extend our result to better fit the common case where
the trace of the Hamiltonian is non-zero (e.g. a dissipative system).

\section{Improved inequality and the marginally passive case\label{sec:Improved and passive}}

While losses are very common in physical systems, gain requires some
special devices, special materials, power sources, etc. Therefore
it is often easier to implement non-unitary evolution in passive systems
(zero gain) at the expense of smaller signals at the output of the
system (e.g. see the discussion in \cite{naimark PT}). As will be
explained next, the basic result (\ref{eq: basic ineq with trace}),
does not provide any useful information for passive systems as the
inequality is trivially satisfied. The goal of this section is to
generalize the previous results and remedy this problem.

Mathematically, a passive system cannot increase the norm of an input
state. At most it can preserve it. In principle there is no point
in attenuating the signal more than needed to reach passiveness. We
shall use the term ``marginally passive evolution'' to describe
a passive system that has at least one state whose norm is unchanged
by $U$, while the norm of other states is either left unchanged or
reduced. By definition the maximal norm of an output vector is given
by the spectral norm of the transformation $U$. The final amplitude
of a state $\ket{\psi_{f}}$ generated by $U$ from a normalized initial
state $\ket{\psi_{i}}$ is simply: $\sqrt{\braket{\psi_{f}}{\psi_{f}}}=\sqrt{\braOket{\psi_{i}}{U^{\dagger}U}{\psi_{i}}}$
. In fact, an alternative definition of the spectral norm is given
by:
\begin{equation}
\max_{\ket{\psi_{i}}}\sqrt{\braOket{\psi_{i}}{U^{\dagger}U}{\psi_{i}}}=\nrm U_{s}
\end{equation}
Therefore, given a general non-unitary $U$, the marginally passive
(MP) evolution operator is given by:
\begin{equation}
U_{MP}=\frac{U}{\nrm U_{s}}.
\end{equation}
Since $\nrm{U_{MP}}_{s}=1$, and more generally for passive evolution
$\nrm{U_{P}}_{s}\le1$, it becomes obvious that (\ref{eq: basic ineq with trace})
is trivially satisfied for any passive evolution (LHS is positive
while the RHS is negative or zero) . In what follows we improve (\ref{eq: basic ineq with trace})
to account for the passive system as well. 

Inspection of (\ref{eq: basic ineq with trace}) reveals that there
is an unnatural asymmetry in the RHS. The ``magnitude'' of the non-unitary
operation is determined by the maximal amplification (maximal singular
value) while the maximal attenuation is completely ignored. Although
in practice amplification is much more difficult to set up in comparison
to attenuation, from a mathematical point of view, they should be
treated on an equal footing. This asymmetry manifests itself clearly
in the MP case. The amplification is zero and all the non-unitary
information is in the degree of attenuation of state. To extract the
attenuation information we will consider the inverse operation $U^{-1}$.
The logic behind doing so, is that amplitude attenuation when going
forward in time is equivalent to amplification when going backwards
in time. We start by taking the time derivative of $U^{-1}U=I$ form
which we get:
\begin{equation}
\partial_{t}U=-U(\partial_{t}U^{-1})U
\end{equation}
upon using it in the Schrödinger Eq. (\ref{eq: schrod U}) we get:
\begin{equation}
-i\partial_{t}U^{-1}=U^{-1}H
\end{equation}
Taking the norm on both sides and using sub-multiplicativity as we
did before, we get:
\begin{equation}
\nrm{\partial_{t}U^{-1}}\le\nrm{U^{-1}}\nrm H
\end{equation}
Repeating the same procedure used to derive (\ref{eq: basic ineq with trace})
we obtain:
\begin{equation}
\int_{0}^{T}\nrm Hdt\ge\ln\left\Vert U^{-1}\right\Vert \label{eq: ineq U^-1}
\end{equation}
Combining it with (\ref{eq: basic ineq with trace}) we get the main
result of this paper:
\begin{equation}
\int_{0}^{T}\nrm Hdt\ge\ln[\mbox{max}\{\nrm U,\left\Vert U^{-1}\right\Vert \}]\label{eq: max U, U^-1}
\end{equation}
Since $\nrm U_{s}\left\Vert U^{-1}\right\Vert _{s}\ge1$ , for passive
systems $\left\Vert U_{P}^{-1}\right\Vert _{s}\ge1/\nrm{U_{P}}_{s}\ge1$.
If so, the attenuation in the system provides a nontrivial result
via the $U^{-1}$ term. Equation (\ref{eq: max U, U^-1}) resolves
the aforementioned asymmetry between gain and loss. Later on it will
be shown more explicitly for the spectral norm. 

There is another way to combine the (\ref{eq: basic ineq with trace})
and (\ref{eq: ineq U^-1}) which is less efficient than (\ref{eq: max U, U^-1})
(i.e. a smaller lower bound), but it offers some insights into the
interplay of gain and loss and a more analytical structure. By adding
(\ref{eq: basic ineq with trace}) and (\ref{eq: ineq U^-1}) and
dividing by two we get:
\begin{equation}
\int_{0}^{T}\nrm Hdt\ge\ln\sqrt{\nrm U\left\Vert U^{-1}\right\Vert }\label{eq: sqrt UU^-1}
\end{equation}
The norm of $U$ in (\ref{eq: basic ineq with trace}) is replaced
by the geometric mean of the norms of $U$ and $U^{-1}$. Moreover
this expression is invariant under changes of the determinant of $U$
(or $\mbox{tr}H$) since $\det U=1/\det U^{-1}$.

The results (\ref{eq: max U, U^-1}) and (\ref{eq: sqrt UU^-1}) apply
to any sub-multiplicative norm. Now we wish to focus on the spectral
norm. It is easy to show that the singular values, $s$, of $U$,
are just the inverse of the singular values of $U^{-1}$ (provided
that $U$ is invertible). Therefore: 
\begin{equation}
\int_{0}^{T}\nrm H_{s}dt\ge\max\{s_{max},s_{min}^{-1}\}\ge\ln\sqrt{s_{max}s_{min}^{-1}}
\end{equation}
where $s_{max}$ and $s_{min}$ are the largest and smallest singular
values of $U$. Here the symmetry between gain and loss appears more
explicitly.

In the marginally passive case $\nrm{U_{MP}}_{s}=1$ so:
\begin{equation}
\int_{0}^{T}\nrm{H_{MP}}_{s}dt\ge\ln\left\Vert U_{MP}^{-1}\right\Vert _{s}\label{eq: ineq MP}
\end{equation}

Finally, to conclude our discussion on the passive case we would like
to point out a general feature of passive systems. It concerns the
tradeoff between ``non-Unitary performance'' and ``detectability''.
As shown in appendix I, $\left|\mbox{det}U\right|{}^{1/N}$ can be
interpreted as the geometric average of the state amplifications generated
by $U$. Starting from $U_{MP}$ and using (\ref{eq: cal U def})
we get:

\begin{eqnarray}
\nrm{\mc U}_{s} & = & \nrm{U_{MP}/\mbox{det}[U_{MP}]{}^{1/N}}_{s}\nonumber \\
 & = & \nrm{U_{MP}}_{s}/\left|\mbox{det}U_{MP}\right|{}^{1/N}\nonumber \\
 & = & 1/\left|\mbox{det}U_{MP}\right|{}^{1/N}
\end{eqnarray}
Combining this with the expression for the geometric mean amplitude,
$\left\langle \left\langle \alpha\right\rangle \right\rangle $, from
appendix I we get:
\[
\left\langle \left\langle \alpha_{MP}\right\rangle \right\rangle =det[U_{MP}]{}^{1/N}=\frac{1}{\nrm{\mc U}_{S}}\le1
\]
This clearly shows the performance vs. detectability tradeoff in the
marginally passive case. The ``stronger'' the non-unitary operation,
is the bigger $\nrm{\mc U}$ is. Yet, as $\nrm{\mc U}$ goes up, the
geometric mean amplification goes down. In other words, if the system
has no gain then stronger non-unitary operation will lead to a weaker
signals.

\section{``Purely non-unitary'' evolution and unitarily invariant norms\label{sec:purely-non-unitary-evolution}}

In \cite{Lidar} a somewhat similar formula to (\ref{eq: basic ineq})
was written for unitary evolution. It follows from \cite{Lidar} that
in the unitary case our $\ln\left\Vert \mathcal{U}(T)\right\Vert $
is replaced by $\left\Vert \ln\mathcal{U}(T)\right\Vert $. Unfortunately
the appealing derivation in \cite{Lidar} is suited only for Hermitian
Hamiltonians and the analysis cannot be easily extended to NH Hamiltonians
and non-unitary evolution. Our derivation overcomes this difficulty
(also up until now we did not require that the norm be unitarily invariant),
at the expense of the capability to capture the resources associated
with unitary evolution. This can actually be an advantage if indeed
our result always captures \textit{only} the resources needed for
the ``purely non-unitary'' part of the evolution. In this section
we clarify in what sense and under what conditions our result (\ref{eq: basic ineq})
quantifies only the purely non-unitary part of the evolution operator.

For simplicity let us assume that $\mathcal{U}(T)$ is diagonalizable
and that $\{\mu_{i}\}$ are the eigenvalues of the evolution operator.
If all eigenvalues satisfy $\left|\mu_{i}\right|=1$, as in the unitary
case, the information on the evolution is encapsulated in $Arg[\mu_{i}]$.
However if the eigenvalues are real and positive (and not all of them
equal to one), it is clear the evolution is ``purely non-unitary''.
The eigenstates are just attenuated or amplified without accumulating
any phase as in the unitary case. With this in mind, we can make use
of the singular value decomposition (SVD). According to the SVD scheme,
any operator $U(T)\in\mathbb{C}^{N\times N}$ (even if it is not diagonalizable)
can be decomposed into \cite{horn 2}:
\begin{equation}
\mathcal{U}(T)=QWP^{\dagger}\label{eq: U SVD}
\end{equation}
where $Q$ and $P$ are some \textit{unitary} matrices%
\footnote{The exact forms of $P$ and $Q$ are not needed for our purpose.%
} and $W$ is a diagonal non-negative matrix whose entries are the
singular values mentioned earlier. In the unitary case $W$ is the
identity matrix. In the non-unitary case $W$ corresponds to a purely
non-unitary evolution operator. Equation (\ref{eq: eig ineq}) shows
that any evolution can be written as a product of unitary evolution
followed by a purely non-unitary evolution and then by another unitary
evolution. Until now we did not require that the norm be unitarily
invariant (UI). However, unitary invariance facilitates a very clear
interpretation of the results. A norm is UI if it satisfies:

\begin{equation}
\nrm A_{UI}=\nrm{CAD}_{UI}
\end{equation}
for any $A\in\mathbb{C}^{N\times N}$ and any unitary $C$ and $D$.
The spectral norm, for example, is unitarily invariant norm. Using
this property and Eq. (\ref{eq: U SVD}) we get :
\begin{equation}
\ln\nrm{\mathcal{U}(T)}_{UI}=\ln\nrm W_{UI}
\end{equation}
That is, the unitary part of evolution is removed and the lower bound
we found addresses only the purely non-unitary part of the evolution.

\section{Examples\label{sec:Examples}}

In this section we demonstrate a few points about the resources inequality
through explicit examples.

\subsection{Example 1 - A simple decay}

Consider a two-level time-independent Hamiltonian with one decaying
level and one stable level: 

\begin{eqnarray}
H & = & \left(\begin{array}{cc}
0 & 0\\
0 & -i\Gamma
\end{array}\right)\\
U & = & \left(\begin{array}{cc}
e^{0} & 0\\
0 & e^{-\Gamma t}
\end{array}\right)
\end{eqnarray}
The traceless Hamiltonian and the corresponding evolution operators
are

\begin{eqnarray}
\mc U & = & \left(\begin{array}{cc}
e^{+\frac{1}{2}\Gamma t} & 0\\
0 & e^{-\frac{1}{2}\Gamma t}
\end{array}\right)\\
\mc H & = & \left(\begin{array}{cc}
+\frac{1}{2}i\Gamma & 0\\
0 & -\frac{1}{2}i\Gamma
\end{array}\right)
\end{eqnarray}
The norm action is:

\begin{equation}
\int_{0}^{T}\nrm{\mc H(t)}_{s}dt=\frac{1}{2}\Gamma T
\end{equation}
The RHS of the resources inequality yield:

\begin{equation}
\ln\nrm{\mc U}_{s}=\frac{1}{2}\Gamma T
\end{equation}
This elementary example demonstrates two important points. The first,
is that the resources inequality (\ref{eq: basic ineq}) can be an
exact equality in some cases. The resources inequality, then, is not
an exaggerated overestimate. The second point is the importance of
the trace removal and the $\mc U$ normalization. For $\Gamma>0$
it is easy to see that $\nrm U_{s}=1$ so $\ln\nrm U_{s}=0$ which
make the resources inequality useless ($\int_{0}^{T}\nrm{H_{0}(t)}_{s}dt\ge0$).
Alternatively, $\Gamma>0$ is exactly the marginally passive case
so instead of normalizing $U$ we can use (\ref{eq: max U, U^-1})
and get:
\begin{equation}
\int_{0}^{T}\nrm{H_{0}(t)}_{s}dt=\Gamma T
\end{equation}

\begin{equation}
\ln\nrm{U_{0}^{-1}}_{s}=\Gamma T
\end{equation}
Which is again an equality rather than an inequality.

\subsection{Example 2 - A cooling evolution}

Consider a quantum two-level system that has two solutions:

\begin{eqnarray}
\ket{\psi_{1}} & = & \left(\begin{array}{c}
1\\
0
\end{array}\right),\label{eq: psi 1}\\
\ket{\psi_{2}} & = & \left(\begin{array}{c}
\sin\frac{\theta(t)}{4}\\
\cos\frac{\theta(t)}{4}
\end{array}\right),\label{eq: psi  2}
\end{eqnarray}
where $\theta(0)=0$. For simplicity we consider $\theta'(t)\ge0$
so that $\theta(t)$ is a monotonously increasing function. While
the two solutions are initially orthogonal to each other, as $\theta\to\pi$
the two states become parallel to each other. This transformation
changes a mixed state density matrix into a coherent one. Alternatively,
this scheme can be used for non-orthogonal state discrimination. If
one starts at $\theta(0)\neq0$ and decreases the angle $\theta$,
two initially non-orthogonal states become orthogonal when $\theta=0$.
Once the states are orthogonal they can be resolved with 100\% certainty.
This kind of non-Hermitian state discrimination was first proposed
in \cite{PT dis}. However, in \cite{PT dis} the needed resources
were not considered. Moreover, in our example the Hamiltonian is time-dependent.
The two solutions (\ref{eq: psi 1}),(\ref{eq: psi  2}) can be used
to construct the evolution operator $U$ from which $\mc U$ and $\mc H$
can be explicitly calculated:

\begin{eqnarray}
\mc H & = & \frac{1}{2}i\theta'(t)\left(\begin{array}{cc}
+\frac{1}{2}\text{Tan}\left[\frac{\theta(t)}{2}\right] & 1\\
0 & -\frac{1}{2}\text{Tan}\left[\frac{\theta(t)}{2}\right]
\end{array}\right)\\
\mc U & = & \frac{1}{\sqrt{\text{Cos}\left[\frac{\theta(t)}{2}\right]}}\left(\begin{array}{cc}
1 & \text{Sin}\left[\frac{\theta(t)}{2}\right]\\
0 & \text{Cos}\left[\frac{\theta(t)}{2}\right]
\end{array}\right)
\end{eqnarray}

\begin{equation}
\nrm{\mc U}_{s}=\sqrt{\frac{1+\sin\frac{\theta}{2}}{\cos\frac{\theta}{2}}}
\end{equation}
Now let us evaluate the norm action. The Hamiltonian norm is:
\begin{equation}
\nrm{\mc H}_{s}=\frac{1}{4}\left|\theta'\right|\frac{1}{\cos[\frac{\theta}{2}]}
\end{equation}
assuming monotonicity ($\theta'\ge0$):
\begin{eqnarray}
\int_{0}^{T}\nrm{\mc H(t)}_{s}dt & =\frac{\theta}{4}+ & \frac{1}{2}\ln\frac{\cos\frac{\theta}{4}+\sin\frac{\theta}{4}}{\cos\frac{\theta}{4}-\sin\frac{\theta}{4}}\label{eq: int H exmp 2}
\end{eqnarray}
After using few trigonometric identities we find that:
\begin{eqnarray}
\ln\nrm{\mc U}_{s}=\frac{1}{2}\ln\frac{1+\sin\frac{\theta}{2}}{\cos\frac{\theta}{2}} & = & \frac{1}{2}\ln\frac{\cos\frac{\theta}{4}+\sin\frac{\theta}{4}}{\cos\frac{\theta}{4}-\sin\frac{\theta}{4}}\label{eq: lnU exmp 2}
\end{eqnarray}
Equations (\ref{eq: int H exmp 2}) and (\ref{eq: lnU exmp 2}) show
that the resources inequality (\ref{eq: basic ineq}) holds ($\theta\ge0$).
In this example the resources inequality was demonstrated for time-dependent
Hamiltonians. 

The resources divergence at $\theta\to\pi$ is a typical feature of
non-unitary cooling and it can be understood without a detailed analysis.
Since at the end of the evolution the two states become almost parallel
to each other, the column vectors of $U$ become linearly dependent.
As a result $\det[U]\to0$. In order to to keep $\det[\mc U]=1$ (see
) the element of $\mc U$ must be very large. The LHS of Eq. (\ref{eq: spec max})
shows that if the elements of an operator are very large, the spectral
norm of the operator is also very large.

The divergence at $\theta\to\pi$ seems to indicate that perfect cooling
is not possible. Nonetheless, the divergence is only logarithmic,
so in practice the norm action does not obtain large values in this
case. For example, if $\theta=\pi-10^{-12}$, the minimal required
resources is only about $15.3$ in units of $\hbar$.

Nonetheless, the divergence is only logarithmic so in practice the
norm action does not obtain large values in this case. For example,
if $\theta=\pi-10^{-12}$ the minimal required resources is only about
$15.3$ in units of $\hbar$. 

The comparison of NH cooling or state discrimination to other methods
(relaxation via the Lindblad equation, POVM) is outside the scope
of this work. Our goal is only to investigate fundamental limitation
on the construction of non-unitary operations generated by the Schrödinger
equation with NH Hamiltonians.

\subsection{Example 3 - Non-Hermitian degeneracy}

In this example we wish to examine the resources in the most extreme
non-Hermitian scenario - the exceptional point Hamiltonian. Exceptional
point Hamiltonians are Hamiltonians that cannot be diagonalized. At
most, similarity transformations can bring these Hamiltonians to the
standard canonical Jordan Block form. See Ch. 9 of \cite{Nimbook}
and references therein for more information on exceptional points.
Consider the following exceptional point Hamiltonian and its corresponding
evolution operator:

\begin{eqnarray}
\mc H & = & \left(\begin{array}{cc}
0 & E_{0}\\
0 & 0
\end{array}\right)\\
\mc U & = & \left(\begin{array}{cc}
1 & -iE_{0}T\\
0 & 1
\end{array}\right)
\end{eqnarray}
The norm action is:

\begin{equation}
\int_{0}^{T}\nrm{H(t)}_{s}dt=\left|E_{0}\right|T
\end{equation}
Since $U^{-1}$ is just $U$ with $E_{0}\to-E_{0}$, the bound based
on $U^{-1}$ gives the same result as the bound based on $U$. The
RHS of the resources inequality is:

\begin{equation}
\ln\nrm U_{s}=\ln\sqrt{1+\frac{1}{2}\left|E_{0}\right|T\,(\left|E_{0}\right|T+\sqrt{4+\left|E_{0}\right|^{2}T^{2}})}
\end{equation}
If the resources inequality holds we should expect: 

\begin{equation}
e^{2\left|E_{0}\right|T}\ge1+\frac{1}{2}\left|E_{0}\right|T\,(\left|E_{0}\right|T+\sqrt{4+\left|E_{0}\right|^{2}T^{2}}),
\end{equation}
Next we wish to show analytically that this inequality indeed holds.

\begin{eqnarray}
e^{2\left|E_{0}\right|T} & = & \sum\frac{(2\left|E_{0}\right|T)^{n}}{n!}\ge1+\left|E_{0}\right|T+(\left|E_{0}\right|T)^{2}\nonumber \\
 & = & 1+\frac{1}{2}\left|E_{0}\right|T(\left|E_{0}\right|T+2+\left|E_{0}\right|T)\nonumber \\
 & \ge & 1+\frac{1}{2}\left|E_{0}\right|T\,(\left|E_{0}\right|T+\sqrt{4+\left|E_{0}\right|^{2}T^{2}})
\end{eqnarray}
This example shows that in some cases the eigenvalues of $\mc U$
are completely inadequate to describe the dynamics. In this example
$\mbox{eig}\{\mc U\}=\{1,1\}$ so the RHS of (\ref{eq: eig ineq})
is equal to zero and therefore the eigenvalues of $\mc U$ carry no
new or useful information on the resources integral. This is similar
to the finding in \cite{100=000025 efficiency}, the evolution speed
near NH degeneracies in the Hamiltonian is determined by the spectral
norm of the Hamiltonian and not by the eigenvalues of the Hamiltonian.
Moreover this example shows that time-independent Hamiltonian does
not necessarily have a norm action which is equal to the bound as
was the case in our first example.

\section{Concluding remarks\label{sec:Concluding-remarks}}

It was shown that non-unitary evolution requires some minimal resources.
The ``norm action'' of the Hamiltonian has to be larger than the
norm of the evolution operator or its inverse. For a passive system
it was also shown that larger non-unitary changes inevitably lead
to weaker signals. \\
The spectral norm of the Hamiltonian was studied before in the context
of evolution speed in non-unitary evolution \cite{100=000025 efficiency}.
Moreover it was used to find the most efficient Hamiltonians for a
given trajectory in the projective Hilbert space. Will these efficient
Hamiltonian also give the minimal norm action (\ref{eq: norm action})?
Not necessarily. The set of constraints in this work is completely
different. Here we consider the final state of all possible initial
states while \cite{100=000025 efficiency} concerns the resources
optimization of a single specific state with a given evolution. Furthermore,
in this work there is no constraint on the evolution between the beginning
and the end of the evolution. \\
Our bound captures only the ``pure non-unitary'' part of the evolution.
In \cite{Lidar} a norm action bound for unitary evolution and Hermitian
Hamiltonians was introduced. It would be interesting to find a bound
that exclusively captures the unitary part of the evolution (for non-unitaty
evolution) as well as a bound that captures everything together and
study the relations between the different bounds.

\ack{}{}

The author wishes to cordially thank Uwe Günther for the suggestion
to investigate the marginally passive case. The generous hospitality
of ITAMP where this work has been completed is greatly acknowledged.

\section*{Appendix I - The geometric mean amplification}

We define the mean amplification in the following way: let $\left\{ \ket m\right\} _{m=1}^{N}$
be an orthogonal set under the standard inner product. The amplification
of each element generated by an evolution operator $U$ is given by:
\begin{equation}
\alpha_{m}=\sqrt{\frac{\braOket m{U^{\dagger}U}m}{\braket mm}}
\end{equation}
The geometric mean is:
\begin{equation}
\alpha_{avg}=\sqrt[_{N}]{\alpha_{1}...\alpha_{N}}=\left(\Pi\sqrt{eig\{U^{\dagger}U\}}\right)^{\nicefrac{1}{N}}=\left|\det U\right|^{\nicefrac{1}{N}}
\end{equation}
To illustrate the reasoning behind the geometric mean consider a two-level
system with two orthogonal eigenstates. One is amplified by a factor
of two and the other by a factor of one-half. The algebraic mean is
equal to $5/4$ while the geometric mean is equal to one. The geometric
mean, then, matches our expectation that the two and one-half should
balance each other. If so, the requirement $\left|\det\mc U\right|=1$
(or equivalently that $\intop_{0}^{t_{f}}\mbox{tr}(H)dt'=0$) used
in section (\ref{sec:A-lower-bound BASIC}), actually means that the
geometric mean amplification is equal to one.

\section*{Appendix II}

Given some time dependent Hamiltonian $H(t)$ the evolution operator
at some time $T$ can be written as:

\begin{equation}
U(T)=\lim_{dt\to0}e^{-iH(T)dt}..e^{-iH(t_{j})dt}...e^{-iH(t_{1})dt}e^{-iH(t_{0})dt},\label{eq: basic U-1}
\end{equation}
where $t_{j}=j\, dt$. Using the sub-multiplicativity and the basic
norm properties we get: 

\begin{equation}
\left\Vert U(T)\right\Vert \le\lim_{dt\to0}\nrm{e^{-iH(T)dt}}...\nrm{e^{-iH(t_{j})dt}}...\nrm{e^{-iH(t_{0})dt}}.
\end{equation}
Moreover, each term satisfies:

\begin{eqnarray}
\nrm{e^{-iH(t_{j})dt}} & = & \nrm{\sum_{k=0}^{\infty}\frac{\left(-iH(t_{j})dt\right)^{k}}{k!}}\le\sum_{k=0}^{\infty}\frac{\left\Vert \left(-iH(t_{j})dt\right)^{k}\right\Vert }{k!}\\
 &  & \le\sum_{k=0}^{\infty}\frac{\left\Vert H(t_{j})\right\Vert ^{k}dt^{k}}{k!}=e^{\nrm{H(t_{j})}dt}
\end{eqnarray}
Therefore:
\begin{equation}
\left\Vert U(T)\right\Vert \le\lim_{dt\to0}e^{\nrm{H(T)}dt}...e^{\nrm{H(t_{j})}dt}...e^{\nrm{H(t_{0})}dt}
\end{equation}
Now the RHS is a multiplication of numbers so:
\begin{equation}
\left\Vert U(T)\right\Vert \le e^{\lim_{dt\to0}\sum_{j}\nrm{H(t_{j})}dt}=e^{\lim_{dt\to0}\sum_{j}\nrm{H(t_{j})}dt}
\end{equation}
From which we once again obtain the resources inequality:
\begin{equation}
\int_{0}^{T}\nrm Hdt\ge\ln\left\Vert U(T)\right\Vert \label{eq: basic ineq-1}
\end{equation}

\section*{Appendix III}

In this appendix, for completeness, we give a simple derivation (\ref{eq: spec of unitary}):
\[
\left\Vert \mathcal{U}(T)\right\Vert _{s}=1\Leftrightarrow\mc{U\mbox{ is unitary}}
\]

One direction is trivial: if $\mc U$ is unitary so that $\mc U^{\dagger}\mc U=I$,
then all the singular values are equal to one. In particular the largest
singular value (the spectral norm) is equal to one. \\
For the other direction we use the fact that $\det\mc U=1$. From
this we get that: 
\begin{equation}
\det\mc U^{\dagger}=(\det\mc U^{T})^{*}=1
\end{equation}
From which follows:
\begin{equation}
\prod s_{i}^{2}=\det\mc U^{\dagger}\mc U=\det\mc U^{\dagger}\det\mc U=1\label{eq: det Udag U}
\end{equation}
where $s_{i}\ge0$ are the singular values of $\mc U$ (the square
root of $\mbox{eig}\{\mc U^{\dagger}\mc U\}$). If $\left\Vert \mathcal{U}(T)\right\Vert _{s}=\max\{s_{i}\}=1$
then to satisfy (\ref{eq: det Udag U}) it must be that all the singular
value are equal to one. Thus, the SVD decomposition of the $\mc U$
is:
\begin{equation}
\mc U=QWP^{\dagger}=QP^{\dagger}
\end{equation}
where Q and P are unitary matrices and therefore:
\begin{equation}
\mc U^{\dagger}\mc U=PQ^{\dagger}QP^{\dagger}=I
\end{equation}
which completes the proof.

\end{document}